%% file: 0-main.tex
\documentclass[reprint,superscriptaddress,aps,prx]{revtex4-2}

\usepackage{graphicx}% Include figure files
\usepackage{dcolumn}% Align table columns on decimal point
\usepackage{bm}% bold math

\usepackage[mathlines]{lineno}% Enable numbering of text and display math
\usepackage{physics}
\usepackage{amssymb}

\def\numx#1e#2{{#1}\mathrm{e}{#2}}

\input{custom-style}

\begin{document}

\preprint{APS/123-QED}

\title{First-principle crosstalk dynamics and Hamiltonian learning via Rabi experiments
 }% Force line breaks with \\

\author{Jan Balewski$^\dag$}
\email{corresponding author balewski@lbl.gov \\$\dag$ equally contributing authors}
\affiliation{
 NERSC, Lawrence Berkeley National Laboratory
Berkeley, CA, USA
}%  balewski@lbl.gov, ORCID: 0000-0002-1899-6526

\author{Adam Winick$^\dag$}
\affiliation{
 Institute for Quantum Computing, Waterloo, ON, Canada
}
\affiliation{
 Q-CTRL, Sydney, NSW Australia \& Los Angeles, CA USA
}

\author{Yilun Xu}
\author{Neel Vora}
\author{Gang Huang}
\affiliation{
ATAP, Lawrence Berkeley National Laboratory
Berkeley, CA, USA\\
}

\author{David Santiago}
\affiliation{
AMCR, Lawrence Berkeley National Laboratory
Berkeley, CA, USA\\
}

\author{Joseph Emerson}  %  jemerson@uwaterloo.ca
\affiliation{
 Institute for Quantum Computing, Waterloo, ON, Canada \\
}

\author{Irfan Siddiqi}
\affiliation{Department of Physics, University of California at Berkeley, Berkeley, CA 94720, USA}
\date{\today}% It is always \today, today,

\begin{abstract} 
Coherent errors constitute a significant barrier to successful large-scale quantum computation. One such error mechanism is crosstalk, which violates spatial locality or the independence of operations.
 We present a description of crosstalk and learn the underlying parameters by executing novel simultaneous Rabi experiments and fitting the Hamiltonian to the observed data. We use this model to predict three- and four-qubit experiments and observe excellent agreement between our theoretical predictions and experimental results. Our technique enables researchers to study the dynamics of multi-qubit circuits without performing experiments, potentially facilitating the minimization of coherent gate errors via digital pulse precompilation. Additionally, this method provides whole-chip crosstalk characterization, a useful tool for guiding quantum processor design.
\end{abstract}

%\keywords{Suggested keywords }%Use showkeys class option if keyword
\maketitle

%\tableofcontents

%%%%%%%%%%%%%%%%%%%%%%%%%%%%%%%%%%%%%%%%%%%%%%%%%%%%%%
%%%%%%%%%%%%%%%%%%%%%%%%%%%%%%%%%%%%%%%%%%%%%%%%%%%%%%

\input{1-intro}
\input{2-theory}

\input{3-experiments}

\input{4-predicting}

\input{5-discussion}
%%%%%%%%%%%%%%%%%%%%%%%%%%%%%%%%%%%%%%%%%%%%%%%%%%%%%%
%%%%%%%%%%%%%%%%%%%%%%%%%%%%%%%%%%%%%%%%%%%%%%%%%%%%%%

\section*{Acknowledgment}
%This research used resources of the National Energy Research Scientific Computing Center (NERSC), a U.S. Department of Energy Office of Science User Facility located at Lawrence Berkeley National Laboratory, operated under Contract No. DE-AC02-05CH11231.
This work was supported by the Office of Advanced Scientific Computing Research, Testbeds for Science program of the U.S. Department of Energy Office of Science under Contract No. DE-AC02-05CH11231 and utilized resources from the National Energy Research Scientific Computing Center (NERSC) at Lawrence Berkeley National Laboratory.

 We would like to acknowledge the fruitful discussions with Akel Hashim, Ravi Naik, Neelay Fruitwala, and Kasra Nowrouzi.

%%%%%%%%%%%%%%%%%%%%%%%%%%%%%%%%%%%%%%%%%%%%%%%%%%%%%%
%%%%%%%%%%%%%%%%%%%%%%%%%%%%%%%%%%%%%%%%%%%%%%%%%%%%%%

\bibliographystyle{apsrev4-2}
\bibliography{references}% Produces the bibliography via BibTeX.

% \fix{Adam, Gang - do you want to thank anyone?}
\appendix
\input{8-append1}

\end{document}

%% file: custom-style.tex
\usepackage{lipsum}
\usepackage{xcolor}  % colored text

\usepackage{qcircuit} % quantum circ
 
\usepackage{circuitikz}

\usepackage{comment} % to exclude large prtions of text

\newcommand{\xt}{\text{crosstalk}}

\usepackage{caption}

%% file: 1-intro.tex
\section{Introduction}

The number of qubits on quantum information processors has rapidly grown over the past decade, and many devices are far too large to simulate exactly with digital computers \cite{Arute2019, Bluvstein2023, Bravyi2024}. Nevertheless, these systems have yet to be used to solve practical problems. A central reason for this apparent disparity is the impact of errors that greatly limit the complexity of realizable quantum circuits.

We can classify errors by their coherence. Decoherent errors are not unitary and thus uncorrectable at the gate level, but we can mitigate their impact with quantum error correction \cite{Shor1995, Gottesman1996, Knill1998, Aharonov1999, Steane2003}. Conversely, coherent errors are unitary and correctable in principle, but characterizing and introducing corrections is exceptionally challenging in practice (they are also more damaging to error thresholds). Crosstalk is a pervasive class of coherent errors that violates one of two assumptions: spatial locality and the independence of operations \cite{Winick2020a,Rudinger2019,Sarovar2019,Abrams2019}. Operations should act on disjoint subsets of qubits. However, unintended interactions can couple qubits or induce errors on non-target subsets.

Ref. \cite{Winick2020a} introduced a scalable framework to model idle and operational crosstalk on experimental devices and showed that, in theory, provided a sufficient degree of control, one can mitigate the effect of crosstalk. In practice, it was unclear how to measure the parameters in the proposed theory and whether such a description of a system is valid.

In this work, we develop a characterization procedure based on novel two-qubit simultaneous Rabi experiments and use the approach to characterize crosstalk on transmon-based processors rigorously. We use the model to predict complex experimental evolutions of up to four simultaneously driven qubits and observe remarkable agreement between our theory and experiments. Across all timescales, the model explains and characterizes virtually all of the formerly elusive coherent errors affecting our superconducting chips.

This paper is structured as follows. First, Sec.~\ref{sec:theory} presents the Hamiltonian that describes classical drive crosstalk. We outline a method for analyzing the model and devise a novel experimental procedure for learning crosstalk parameters. Next, in Sec. \ref{sec:experiments}, we execute the procedure on a real 8-qubits, fixed-frequency transmon-based quantum processor to fully characterize the classical driving fields crosstalk. 
%We crosscheck against alternative methods and see consistent results and much higher precision than was formerly possible. 
In Sec.~\ref{sec:predicting}, we conduct verification experiments where we simultaneously drive three and four qubits. We compare the theory's predictions and observed dynamics using linear superposition of the two-qubit pairwise information extracted in Sec.~\ref{sec:experiments}. We find a good  agreement, which provides strong evidence that the model is correct. 
Finally,  in Sec.~\ref{sec:discussion} we discuss practical applications of this tool for understanding and mitigating crosstalk errors, including  mitigating crosstalk through control optimization.

%% file: 2-theory.tex
\section{Theory of crosstalk} \label{sec:theory}

We begin by analyzing the Hamiltonian corresponding to a transmon~\cite{Koch2007} subject to an arbitrary collection of semiclassical driving fields \cite{Gambetta2017,Krantz2019}. Ideally, in this system, drive $j$ should only affect transmon $j$ and no other qubits, ensuring precise control without crosstalk. However, due to practical limitations, some crosstalk between drives and qubits is inevitable. The semiclassical Hamiltonian corresponding to transmon $j$ with local drives crosstalk due to $k$ other dives  is  
\begin{equation}
H^{(j)} = H_0^{(j)} + \sum_k H_d^{(jk)} \,,
\label{eq:genHam}
\end{equation}
where the transmon's Hamiltonian, $H_0^{(j)}$, and drive Hamitonians, $H_d^{(jk)}$, are
\begin{align}
H_0^{(j)} &= \omega_j\hat n + \frac{\alpha_j}{2}\hat n(\hat n - 1) \,, \\
H_d^{(jk)} &= \beta_{jk}\Omega_k(t_{jk})\cos(\omega_k't_{jk}-\theta_{jk}-\phi_k)(\hat a + \hat a^\dagger) \,.
\label{eq:2qHam}
\end{align}

The Hamiltonian $H_0^{(j)}$ is an approximate model describing the low-energy states of transmon $j$, which behaves as a Duffing oscillator. The variables $\omega_j$ and $\alpha_j$ denote the transmon's frequency and anharmonicity, respectively. The driving Hamiltonian $H_d^{(jk)}$ is the effect induced  on qubit $j$ by driving field $k$. The variable $\beta$ specifies the relative \xt\ strength ($\beta_{jk}\geq 0$), $\theta$ is the relative \xt\ phase, and $\phi_k$ is a phase shift set by software \cite{McKay2016}. By convention, when there is a one-to-one correspondence between drives and transmons, we should have $\beta_{jj}\approx1$ and $\theta_{jj} = 0$. This imposes the notion that we intend drive $j$ to control transmon $j$. We are free to fix $\theta_{jj}=0$ since this specifies the measurement basis. However, since perfect calibration is  hard to achieve, $\beta_{jj}$ is not necessarily exactly equal to 1. The function $\Omega_k$ is the pulse envelope, and $\omega_k'$ is the drive frequency. Since the time for a driving field to reach a qubit is not the same for all qubits, i.e., the signal path lengths differ, we introduce $\tau_{jk}=\tau_j-\tau_k$, which captures a relative time delay. The matrix $\tau$ is always skew-symmetric. Moreover, any column or row of $\tau$ completely specifies the rest of the matrix. We define $t_{jk}=t-\tau_{jk}$ to simplify the presentation.
%\jan{How about put $t_j-\tau_{jk}$ into eq. 3 and not define $ t_{jk}$ at all?}

To focus on the problem of measuring crosstalk, we consider a system with two driving fields and two transmons, labeled $a$ and $b$. Each field targets one transmon, but crosstalk introduces unintended effects on the other. The crosstalk parameters $\beta_{jk}$, $\theta_{jk}$, and time delays $\tau_{jk}$ reduce to a $2 \times 2$ matrix, with $\tau_{jk} = -\tau_{jk}$. This simplification allows a clear analysis of pair-wise cross-talk and its impact on quantum control. We denote measured qubit as $a$  ($j=a$) and the other qubit as $b$ ($k=b$). Then,
%In the case where there are exactly two driving fields $a$ and $b$,
$t_{ab}=t$, and assuming we can neglect the anharmonic behavior of the systems and $\beta_{aa}=1$, we can analytically solve the system \eqref{eq:2qHam} and obtain
\begin{align}
\expval{X}_{Qa;Qb} &= \left[\sin\phi_a + \beta_{ab}\sin(\theta_{ab}+\phi_b) \right] \frac{\sin \eta_{ab}\expval{\Omega}t}{\eta_{ab}} \,, \\
\expval{Y}_{a;b} &= \left[\cos\phi_a + \beta_{ab}\cos(\theta_{ab}+\phi_b) \right] \frac{\sin \eta_{ab}\expval{\Omega}t}{\eta_{ab}} \,, \\
\expval{Z}_{a;b} &= \cos \eta_{ab}\expval{\Omega} t \,,
\label{eq:evZ}
\end{align}
where
\begin{equation}
\eta_{ab} = \sqrt{1+\beta_{ab}^2+2\beta_{ab}\cos(\Delta\phi_{ab}-\theta_{ab})} \,,
\end{equation}
In the limit of weak \xt\
\begin{equation}
\eta_{ab} \approx \left(1+\beta_{ab}\cos(\Delta\phi_{ab}-\theta_{ab})\right) \,,
\end{equation}
and if we fix drive duration $t$ so $\hat\Omega=\expval{\Omega}t = \pi/2 + n\pi$
\begin{equation}
\expval{Z}_{a;b} \approx \sin(\beta_{ab}\hat\Omega\cos(\Delta\phi_{ab}-\theta_{ab})) \,.
\end{equation}

%% file: 3-experiments.tex
\section{Learning \xt\ experimentally} \label{sec:experiments}

\begin{figure}[htbp]
\scalebox{0.99}{
\Qcircuit @C=0.7em @R=0.7em {
&&\lstick{\ket{0}_{a}} & \qw&\qw&\gate{R_x(\hat\Omega)} &\qw& \measuretab{\mbox{Z}}   \\
&& \lstick{\ket{0}_{b}} & & \gate{R_z(\Delta\phi)} &\gate{R_x^{f_a}(\hat\Omega)} & \qw   \\
}
}
\caption{ \raggedright Quantum circuit for the 2-qubits experiment used to measure cross-talk induced on  qubit $a$ due to simultaneous drive  $b$. }
 \label{qc:exp2}
\end{figure}
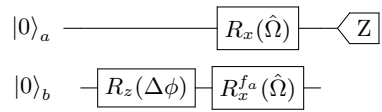

We designed a  protocol to measure the directional \xt\ between two qubits using  circuit shown in Fig.~\ref{qc:exp2}. 
A standard $R_x(\hat\Omega)$ rotation is applied to the `primary' qubit $a$, characterized by the driving frequency $f_a$.  The symbol \raisebox{0.25em}{\Qcircuit @C=0.2em {& \qw & \measuretab{Z}}} denotes measuring the expectation value (EV) of this qubit in the $Z$-basis. 
Simultaneously, a non-standard pulse is applied to the other qubit $b$ as a modified $R_x$ rotation, with its driving frequency shifted from $f_b$ to $f_a$. To enhance the \xt\ signal and reduce the shot cost, we use a relatively large rotation angle $\hat\Omega = 2.5\pi$ for both $R_x$ gates, inducing a longer interaction time $t=160$ ns between the two qubits.

% %%%%%%%%%%%%%%%%%%%%%%%%%%%%%%%%%%%%%%%%

For a given pair of qubits, we perform a series of measurements, varying the  phase of the virtual $R_z(\Delta\phi)$ rotation applied to drive $b$ in the range $\in [0, 2\pi]$. According to our \xt\ model, the  $\Delta\phi$ dependence of the expectation values measured in the z-basis is  described by Eq.~\eqref{eq:evZ}, with  $\Delta\phi=\Delta\phi_{ab}$.
Since the \xt\ can vary for each directional qubits pair, separate measurements must be performed for each pair.

The experiments were performed on a superconducting quantum processor 
\cite{hashim2023quasi,hashim2024efficient}
fabricated and operated by the Advanced Quantum Testbed (AQT) at LBNL  \cite{AQTwebsite}. 

\begin{figure}[htbp]
\centering
\begin{tikzpicture}[node distance={15mm}, main/.style = {draw, circle, minimum size=6mm, inner sep=0.5mm}] 
\node[main] (0)              { $0$}; 
\node[main] (1) [right of=0] {$1$};
\node[main] (2) [right of=1] {$2$};
\node[main] (3) [right of=2] {$3$};
\node[main] (7) [below  of=0] {$7$};
\node[main] (6) [right of=7] {$6$};
\node[main] (5) [right of=6] {$5$};
\node[main] (4) [right of=5] {$4$};
\draw (0) -- (1);
\draw (2) -- (1);
\draw (2) -- (3);
\draw (4) -- (3);
\draw (4) -- (5);
\draw (6) -- (5);
\draw (6) -- (7);
\draw (0) -- (7);
\end{tikzpicture}
  \caption{ Schematic native connectivity for the AQT chip  used for testing of the \xt\ modeling. 
  }
\label{fig:chipTopo}
\end{figure}
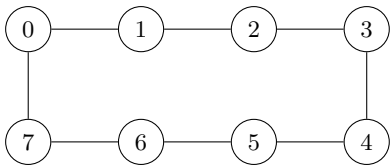

The processor has a 8-qubit ring topology, with resonator-mediated couplings between nearest neighbors. The schematic connectivity of this chip is shown in Fig.~\ref{fig:chipTopo}. The image of the physical chip can be found in \cite{hashim2023quasi}, Fig. 1a.

%=================================================

In this work, we opted for simple cosine-shaped gate envelopes instead of using DRAG~\footnote{Derivative Removal by Adiabatic Gate} \cite{Motzoi2009}. Given the relatively long duration of our 160 ns pulses, and anharmonicity exceeding 600 MHz, the DRAG correction would be negligible. Moreover, our characterization is based on the total area of pulses, and the integral of DRAG corrections is zero.

For every pair of qubits we executed  circuit shown in Fig.~\ref{qc:exp2} for 33 uniformly distributed values of $\Delta\phi$, taking 1000 shots per data point. The readout error was corrected in post processing using the {\it M3-method}~\cite{M3SPAM}.

At the time of executing experiments on the chip,  qubit $5$ operated incorrectly (or its calibration drifted away too quickly) hence we discarded all data when $5$ played the role  readout qubit $a$. But the electronics driving $5$ was functioning correctly so $5$ drive could still be used as  a valid source of the \xt\ for other qubits.  The \xt\ results for the remaining 49 qubit pairs are  discussed in the following material.

\begin{figure}[htbp]
\centering
\includegraphics[width=0.99\linewidth]{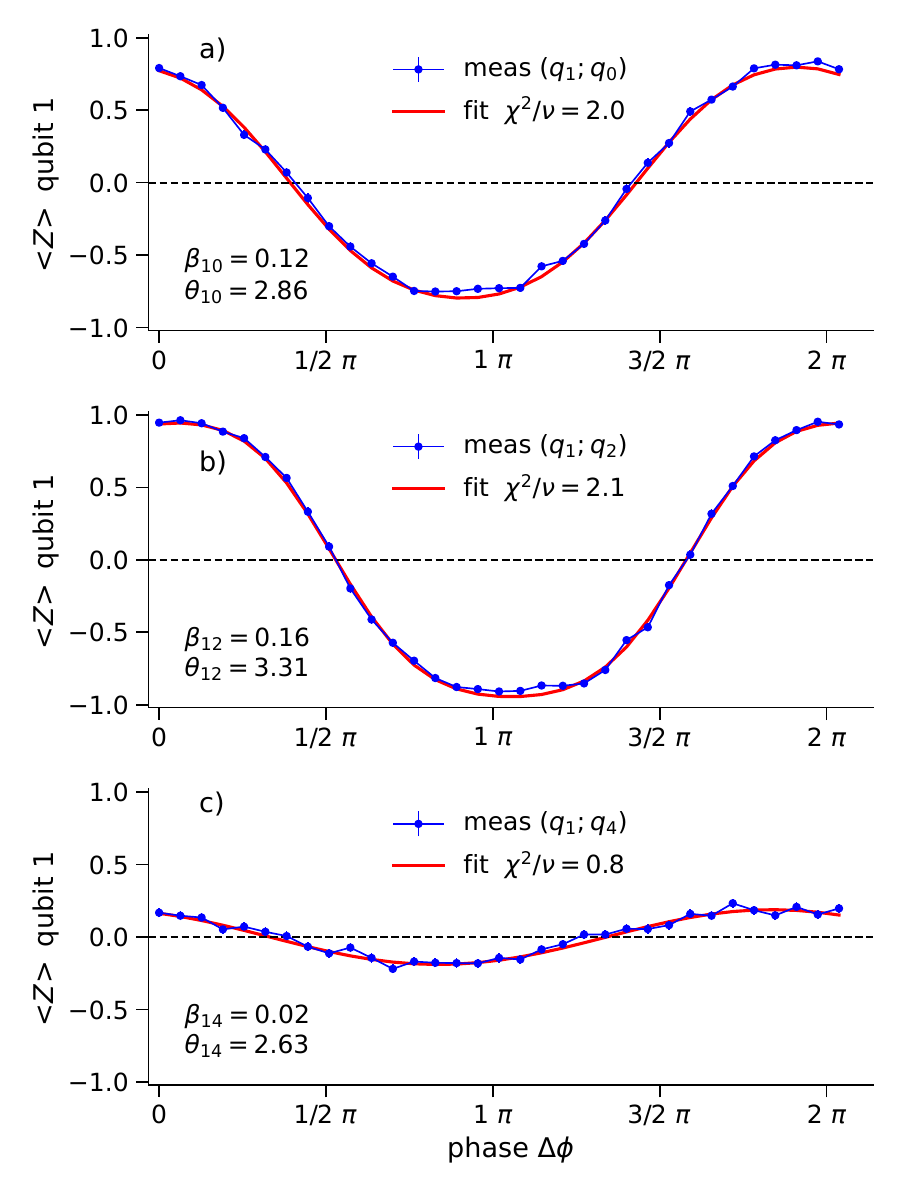}
   \caption{Examples of measured \xt\  using circuit shown in Fig.~\ref{qc:exp2}.  The  expectation value for $\sigma_z$ for qubit 1 is on the y-axis. We  simultaneously drive one other qubit $\{0,2,4\}$, shown in panels a)–c), respectively. The varied phase $\Delta\phi$ is plotted on the x-axis.   The fitted \xt\ model, Eq.~\eqref{eq:evZ}, is shown as red line, its strength $\beta_{ab}$ and phase $\theta_{ab}$ are displayed in the lower-left corner.
}
\label{fig:xtFit}
\end{figure}

Example of measured expectation values for qubit 1 as function of phase  $\Delta\phi$ applied to one of three other qubits: $b\in \{0,2,4\}$ are shown in Fig.~\ref{fig:xtFit}.
The \xt\ model, given by Eq.~\eqref{eq:evZ}, has two fitted parameters: strength $\beta$ and phase $\theta$. The red line depicts fitted model which matches well each of 3 distributions. As an objective function we used chi-squared per degree of freedom ($\chi^2/\nu$) which is a statistical measure used to assess the goodness of fit between observed data and a theoretical model~\cite{Bevington2003Data}.
\begin{align}
\chi^2/\nu &= \frac{1}{N - p}~\sum_{i=1}^{N} \frac{\left(O_i - M_i\right)^2}{\sigma_i^2} 
\label{eq:chi2}
\end{align}
where  \(M_i\) is the model value for the \(i\)-th data point, \(O_i\) and \(\sigma_i\)  are  the observed values and their  standard deviation, \(N\) is the total number of data points, and \(p=2\) is the number of model parameters.

The values of $\beta,\theta$, and $\chi^2/\nu$ are also shown on each panel of Fig.~\ref{fig:xtFit}.
The \xt\ is stronger $\beta>0.1$ for 2 qubit pairs: 1-0 and 1-2 which also share a coupler.

We note, that for an ideal qubits the \xt\ should not exist ($\beta$=0) and measured distributions for circuit shown in Fig.~\ref{qc:exp2}  should be perfectly flat, y=0.5. 

% JAN: DO NOT REMOVE
% /xtalkmitigation/2024_paper/$ ./figA_exp2.py
%=================================================

\begin{figure}[htbp]
\centering
\includegraphics[width=0.99\linewidth]{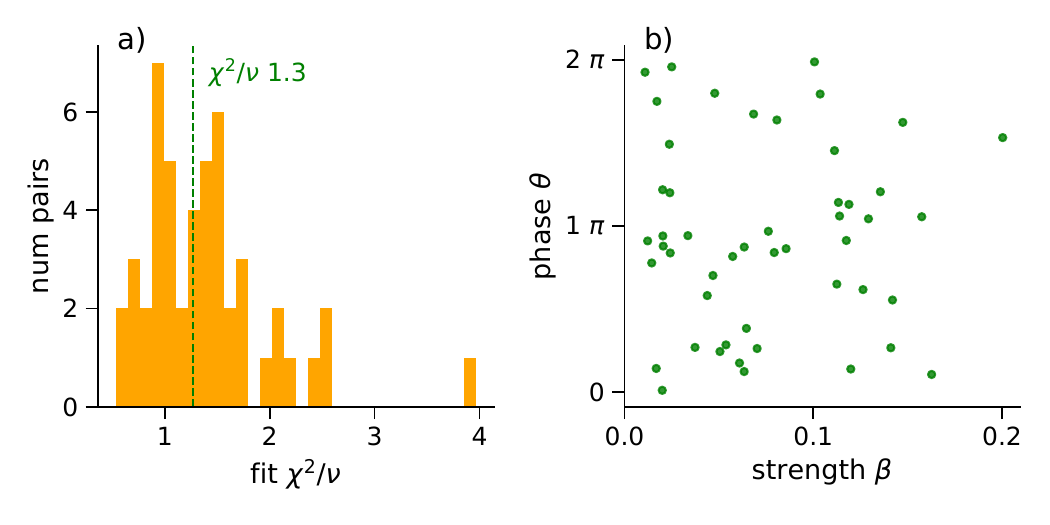}
   \caption{Statistical characterization of \xt\  strength $\beta$ and phase $\theta$ measured for the chip using model in Eq.~\eqref{eq:evZ}. a) Distribution of $\chi^2/\nu$ , green vertical line denotes median. 
   b) There is no correlation  between  $\beta$ and $\theta$ for the investigated chip.
}
\label{fig:xtStats}
\end{figure}

We measured \xt\ for all 49 functioning qubit pairs on the chip. The distribution  of $\chi^2/\nu$ is shown in Fig.~\ref{fig:xtStats}a.
The median value of $\chi^2/\nu$ of approximately 1 indicates that the discrepancies between the observed data and the fitted model  are consistent with the expected random statistical fluctuations, signifying a good fit.
No statistically significant correlation was observed between  strength $\beta$ and phase $\theta$, shown in Fig.~\ref{fig:xtStats}b.

% JAN: DO NOT REMOVE
% /xtalkmitigation/2024_paper/$  ./figB_xtalkFitStats.py 
%=================================================

\begin{figure}[htbp]
\centering
\includegraphics[width=0.7\linewidth]{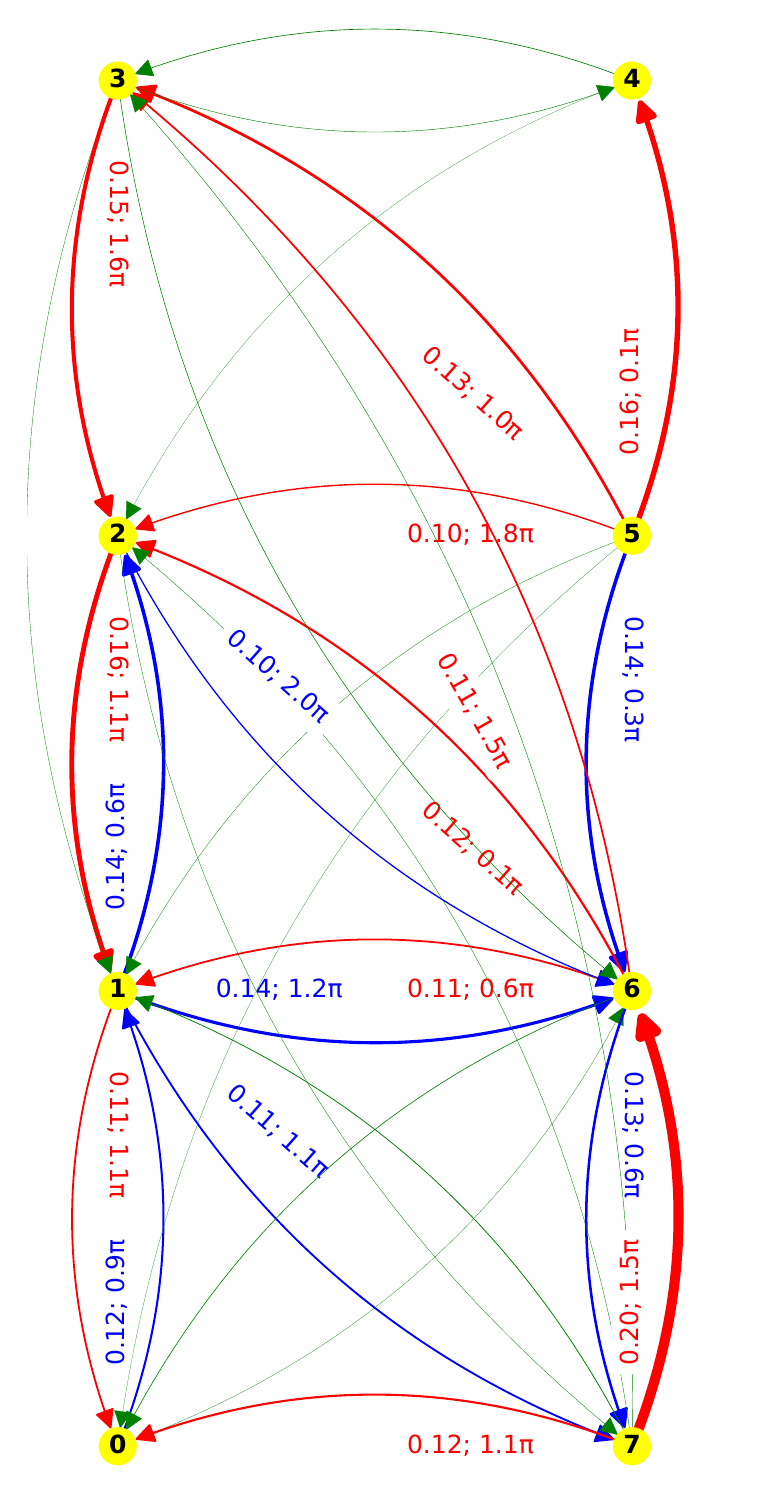}
   \caption{Graphical representation of the measured \xt\ strength, $\beta$, between all qubits pairs on the chip. The edges with  $\beta>0.1$ are shown in red or blue depending on directionality.   The edges  with small $\beta<0.05$ are not shown, other edges with intermediate $\beta$ are shown in green. The thickness of edges is correlated with magnitude of $\beta$.  The  printed values  denote \xt\ $(\beta,\theta)$ of red or blue edges. 
}
\label{fig:xtGraph}
\end{figure}

 Graphical representation of  measured strengths of the \xt\  between all qubits pairs on the chip   is shown in Fig.~\ref{fig:xtGraph}. The thickness of arrows is correlated with magnitude of $\beta$ and only edges with   $\beta>0.05$ are shown.
We observe that qubits 1,2, and 6 seem to be susceptible to \xt\  vs. most other qubits. Also \xt\ seems be to stronger when a physical couplers exist (see connectivity in Fig.~\ref{fig:chipTopo}). For a given pair of qubits \xt\ is often stronger in one direction vs. the opposite one. 
A potential explanation for the significant \xt\ observed between spatially separated qubits on the chip could be the proximity of a qubit's frequency to a {\it box mode frequency}. This proximity might induce unintended coupling between the qubit and the mode, possibly leading to energy leakage and degraded qubit performance~\cite{Bronn_2018}.

% JAN: DO NOT REMOVE
% /xtalkmitigation/2024_paper/$ ./figC_xtalkGraph.py 
%=================================================

%% file: 4-predicting.tex
\section{Predicting \xt\ }\label{sec:predicting}
So far we have discussed the procedure of extracting   \xt\ strength $\beta$ and phase $\theta$ for each pair of physical qubits to inform   the 2 qubit  Hamiltonian model given by Eq.~\eqref{eq:2qHam}.  We have measured $\beta_{ab},\theta_{ab}$ values for all working qubit pairs a-b on the chip.  Since general  Hamiltonian, Eq.~\eqref{eq:genHam}, is the sum of 2 qubits \xt\ terms we can now construct   Hamiltonian describing \xt\  for  the 3 or more qubits circuit
  without additional measurements, by simply adding relevant terms. 

\begin{figure}[htbp]
\centering
\scalebox{0.95}{
\Qcircuit @C=0.7em @R=0.7em {
&&\lstick{\ket{0}_{a}} & \qw&\qw&\gate{R_x(\hat\Omega)} &\measuretab{\mbox{Z}}   \\
&& \lstick{\ket{0}_{b}} & & \gate{R_z(\Delta\phi)} &\gate{R_x^{f_a}(\hat\Omega)} & \qw   \\
&& \lstick{\ket{0}_{c}} & & \gate{R_z(\Delta\phi)} &\gate{R_x^{f_a}(\hat\Omega)} & \qw   \\
}}
\caption{Quantum circuit for  3 qubits experiment used to verify  learned cross-talk model.}
 \label{qc:exp3}
\end{figure}
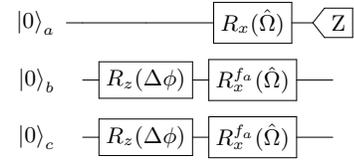

The 3-qubits circuit shown in Fig.~\ref{qc:exp3} used for \xt\ verification is very similar to the  Fig~\ref{qc:exp2}. The drive on the 3rd qubit is again the modified $R_x$ gate with its driving frequency shifted from $f_c$ to $f_a$.

The predicted 3-qubit \xt\ model $\expval{Z}_{a;bc}$ is expressed as follows:
\begin{align}
    \eta_{a;bc}^2 &= 1 + \beta_{ab}^2 + \beta_{ac}^2 + 2\beta_{ab}\cos(\Delta\phi-\theta_{ab}) \nonumber \\
    &+ 2\beta_{ac}\cos(\Delta\phi-\theta_{ac}) \nonumber\\
    &+ 2\beta_{ab}\beta_{ac}\cos(\theta_{ab} - \theta_{ac}) \\
    \expval{Z}_{a;bc}&=\cos( \hat\Omega \cdot \eta_{a;bc} )
\end{align}

We have also taken data for the 4-qubits version of the circuit, which we do not show because it looks too similar.
The predicted 4-qubit model $\expval{Z}_{a;bcd}$ is:
\begin{align}
\eta_{a;bcd}^2 &= 1 + \beta_{ab}^2 + \beta_{ac}^2 + \beta_{ad}^2   \\
    &+ 2\beta_{ab}\cos(\Delta\phi-\theta_{ab}) 
    + 2\beta_{ac}\cos(\Delta\phi-\theta_{ac}) \nonumber\\
    &  + 2\beta_{ac}\cos(\Delta\phi-\theta_{ad}) 
    + 2\beta_{ab}\beta_{ac}\cos(\theta_{ab} - \theta_{ac}) \nonumber\\
    &+ 2\beta_{ab}\beta_{ad}\cos(\theta_{ab} - \theta_{ad}) 
     + 2\beta_{ac}\beta_{ad}\cos(\theta_{ac} - \theta_{ad}) \nonumber\\
    \expval{Z}_{a;bcd}&=\cos(\hat\Omega\cdot \eta_{a;bcd})  
    \label{eq:predExp3z}
\end{align}

% %%%%%%%%%%%%%%%%%%%%%%%%%%%%%%%%%%%%%%%%

% %%%%%%%%%%%%%%%%%%%%%%%%%%%%%%%%%%%%%%%%

Since the number of possible choices of 3 or 4 qubits out of 8 is very large, we limited experimental verification to about 40 qubits multiplets,  chosen at random.  Similarly, for each mutiplet we  collected data at 33 uniformly distributed values of $\Delta\phi$, taking 1000 shots per data point, corrected off-line for the readout error using the M3-method.

\begin{figure}[htbp]
\centering
\includegraphics[width=0.99\linewidth]{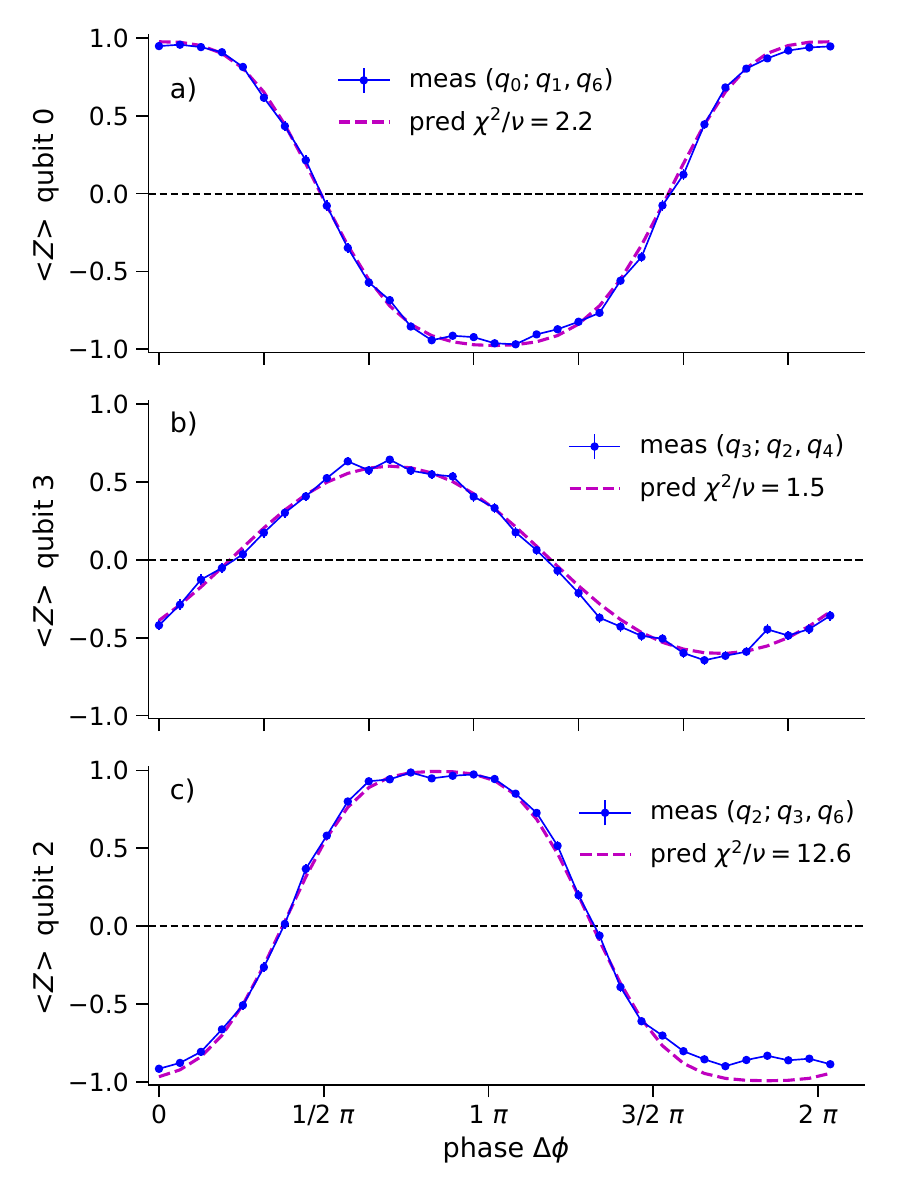}
   \caption{Examples of measured 3-qubits \xt\  using circuit shown in Fig.~\ref{qc:exp3}.  The  expectation value for $\sigma_z$ for qubit $a$ is on the y-axis.   The predicted \xt\ model using Eq.~\eqref{eq:predExp3z} is shown as magenta dashed lines. Printed $\chi^2/\nu$  measures amount of disagreement between the data and predictions.
}
\label{fig:predExp3z}
\end{figure}

Examples of measured and predicted \xt\ for selected 3 qubit experiments are shown in Fig.~\ref{fig:predExp3z}. Panels a) and b) show typical good match between experiment and prediction with  $\chi^2/\nu$ close to 1. About half of the measured triplets were accurately predicted. Panel c) shows a case where the prediction missed the measured distribution as the EV approached -1. We attribute this to the less effective cosine-envelope used for the \( R_x \)-gate, which could result in the trajectory never reaching the \( \ket{1} \) state due to uncorrected higher harmonics. Independently, qubit 2 may have experienced instability potentially caused by environmental two-level systems (TLS), evidence for which is beyond the scope of this paper. Although  computed $\chi^2/\nu=13$ is very large,  the prediction captured the experiment very well over 2/3 of the $\Delta\phi$ range.

% JAN: DO NOT REMOVE
% /xtalkmitigation/2024_paper/$  ./figD_exp34.py --expType exp3z
%=================================================

\begin{figure}[htbp]
\centering
\includegraphics[width=0.99\linewidth]{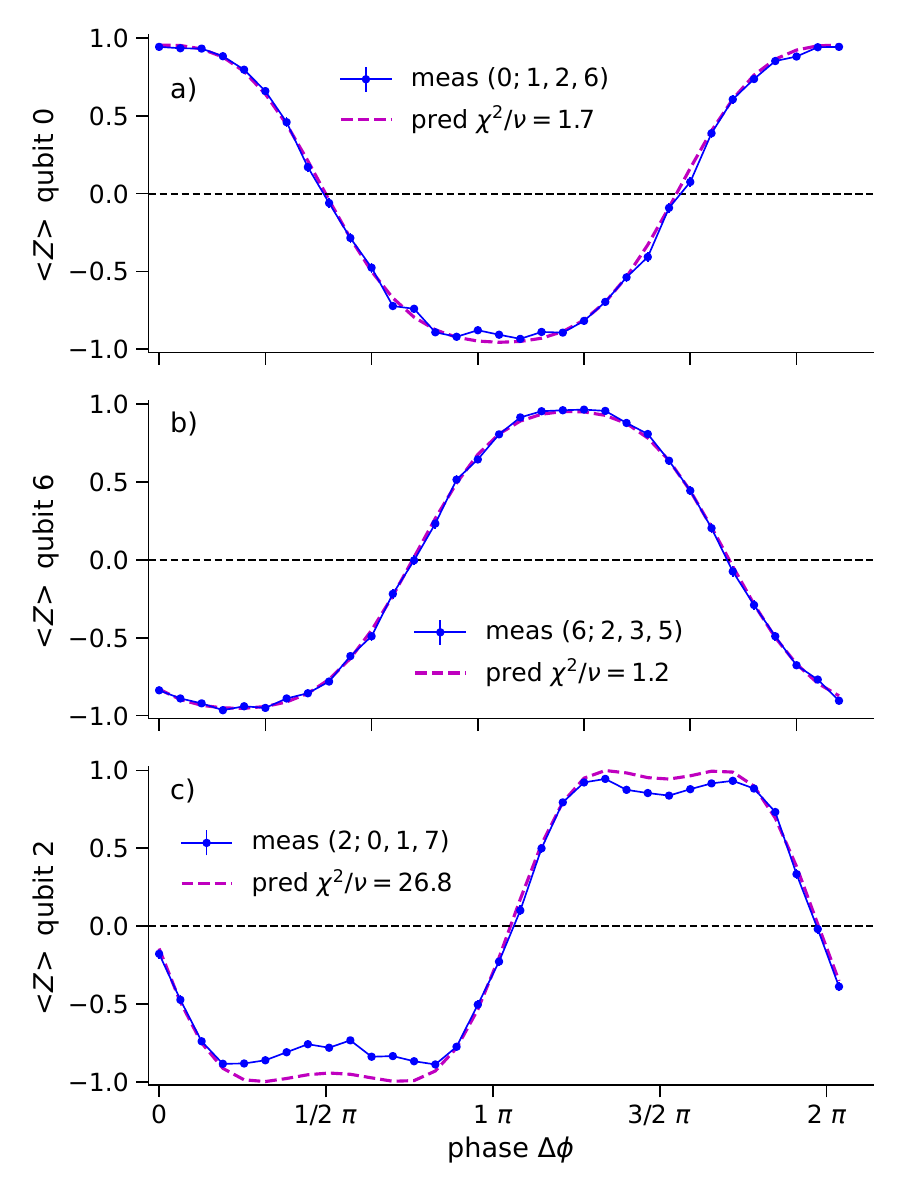}
   \caption{Examples of measured and predicted \xt\  for 4 qubit experiments. See caption of  Fig.~\ref{fig:predExp3z} for the details.
}
\label{fig:predExp4z}
\end{figure}

Examples of measured and predicted \xt\ for selected 4 qubit experiments are shown in Fig.~\ref{fig:predExp4z}. Panels a) and b) show typical good match. Panel c) shows a case for qubit 2 when prediction deviates as measured  EV approaches 1 or -1, most likely caused by effects discussed earlier. 

It should be noted, in the absence of the \xt\, i.e. for an ideal chip and drives,  the data on Figs.~\ref{fig:predExp3z},\ref{fig:predExp4z} should show flat distributions at y=0. We run experiments on a  chip which had a relatively large \xt\ up to 15\%. Our model was able to predict most of the \xt\ induced distortions, the residual between the blue data  and dashed magenta lines are rather small.

% JAN: DO NOT REMOVE
% /xtalkmitigation/2024_paper/$  ./figD_exp34.py --expType exp4z
%=================================================

\begin{figure}[htbp]
\centering
\includegraphics[width=0.99\linewidth]{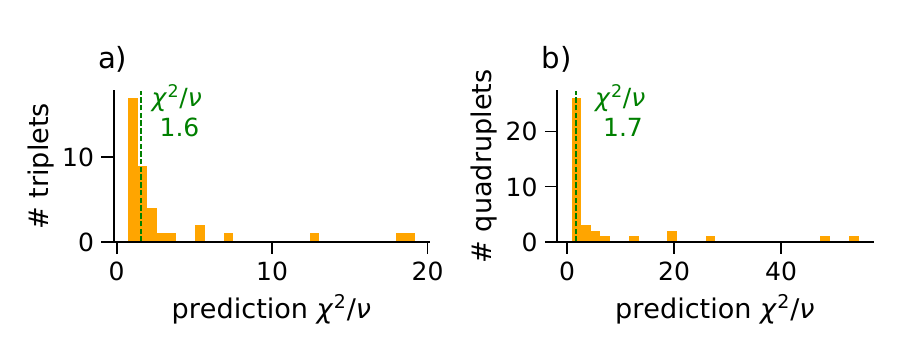}
   \caption{Distribution of $\chi^2/\nu$  for \xt\ prediction for randomly chosen  (a) triplets  and (b)  quadruplets  of qubits. Green  vertical lines denote the median. 
}
\label{fig:predStat}
\end{figure}

To summarize the quality of the  \xt\ prediction for about 80 randomly chosen  multiplets of qubits we plotted histograms of  $\chi^2/\nu$, 
shown in  Fig.~\ref{fig:predStat}.
The median of $\chi^2/\nu$ is below 2, meaning our Hamiltonian  model reasonably predicts 3- and 4-qubits \xt\ experiments with accuracy of about 4\% for most of the cases.

We took 1000 shots per data point which reduces the statistical error of experiments to about 3\%, entering in the denominator in Eq.~\eqref{eq:chi2}. In this context one should interpret 'large' $\chi^2/\nu$  as discrepancies exceeding few \%, which is still very promising agreement.

% JAN: DO NOT REMOVE
% /xtalkmitigation/2024_paper/$  ./figE_xtalkPredStats.py 
%=================================================

%% file: 5-discussion.tex
\section{Discussion} \label{sec:discussion}

We have developed a practical protocol for whole-chip crosstalk characterization, enabling comprehensive noise analysis of quantum processors. By executing novel two-qubit Rabi-like measurements between all pairs of qubits, our method efficiently determines the crosstalk parameters across the entire chip. For a chip with all-to-all crosstalk our model requires \( O(N(N - 1)) \) experiments. In practice, larger devices have some degree of locality, so the total number of required experiments is \( O(N) \) and can possible be further be reduced to \( O(1) \) if we can apply operations to different regions of a device simultaneously. This scalability makes our approach practical for larger quantum processors.

The Hamiltonian modeling crosstalk is linear and once it is primed with 2-qubits data it is capable of predicting crosstalk distortions for circuits with an arbitrary number of qubits. The model's ability to predict crosstalk effects in larger circuits is a significant advantage for studying and mitigating errors in quantum computations.

We demonstrated experimentally that crosstalk can be learned for an 8-qubit chip provided by AQT at LBNL. Our Hamiltonian model reasonably predicts three- and four-qubit experiments with an accuracy of about 4\% for most cases. This level of precision indicates that the bulk of the distortions caused by crosstalk are captured by our model, validating its effectiveness in practical settings.

For some cases, the predicted crosstalk deviates from the experimental results. We attribute this discrepancy to the less accurate cosine envelope used for the \( R_x \)-gate. Additionally, one of qubits  might have experienced instability caused by an unstable environmental two-level system (TLS). 

This work sets the mathematical foundations for correcting two-qubit pulses to mitigate crosstalk. While pulse correction is beyond the scope of this work, our model provides a basis for future development of pulse optimization techniques that can reduce coherent errors due to crosstalk. Implementing such corrections could significantly enhance the fidelity of quantum gates in noisy environments.

In comparison to other crosstalk characterization methods like Randomized Benchmarking (RB), our method requires much less prior calibration of the chip. We require for each qubit only a reasonable \( R_x \) gate and knowledge of the readout error. The calibration circuits we execute are very shallow, making our method less sensitive to decoherence times. However, we require pulse-level access to change the frequency of the gate drive, which may not be available to end-user on all quantum hardware platforms.

In the future work the Derivative Removal by Adiabatic Gate (DRAG) envelope should be used, which requires no  modification in the Hamiltonian model and formulas used in this paper. This adjustment is essential for accurately modeling systems that employ advanced pulse-shaping techniques.

Our approach provides a scalable and efficient method for characterizing crosstalk in multi-qubit systems. By understanding and modeling crosstalk effects, researchers can better predict the behavior of quantum circuits and design strategies to mitigate errors. This work contributes to the broader goal of improving quantum processor performance and advancing towards large-scale quantum computation. Future research could focus on integrating pulse correction techniques and extending the model to include higher-order interactions, further enhancing the reliability of quantum computations.

%% file: 8-append1.tex
%\clearpage % Ensure a new page for the appendix
\begin{widetext}

\section{QubiC - quantum processor controller}

% ......  QubiC description - perhaps 30% to long, keep as-is for now
Experimental data for this work were collected using 
Qubit Control ({\it  QubiC })   system \cite{xu2021qubic,xu2023qubic}. QubiC is a field-programmable gate array (FPGA)-based, open-source, full-stack control system designed for quantum processors. 
It generates and routes complex sequences of radio frequency (RF) signals from control electronics for qubit pulse generation and quantum state measurement. 
Leveraging advances in radio frequency system-on-chip (RFSoC) technology, QubiC uses the AMD ZCU216 evaluation board and incorporates a range of enhanced features. 
The system employs portable FPGA gateware and a streamlined processor to handle real-time command execution.
QubiC utilizes a multi-core distributed architecture, assigning a dedicated processor core to each qubit. 
The system generates pulses by combining predefined pulse envelopes and carrier information as specified in the command. 
These pulse envelopes are pre-stored in the FPGA's block random-access memories (RAMs), ensuring rapid execution and reusability across the quantum circuit. 
Pulse parameters such as amplitude, phase, and frequency can be dynamically updated for each pulse.
The QubiC software stack compiles quantum programs into binary commands, which are then loaded into the FPGA for execution.

\section{Accumulation of \xt}
It is instructive to trace how \xt\ predicted for experiment with 4 concurrent drives occurs.
We selected several measurements when we always measure qubit 0 as the primary qubit.  Fig.~\ref{fig:xtOrigin}a shows {\bf predictions} for combined \xt\ due to driving qubit 0  and simultaneously 3 other qubits: 1,2, and 6.
 Fig.~\ref{fig:xtOrigin}b,c are also showing {\bf predictions}  for  3-qubits \xt\ due secondary qubits being 1,6 or  2,6.
 Fig.~\ref{fig:xtOrigin}d-f show {\bf fits} of \xt\ for 
 individual pairs. We note  \xt\  due to qubit 2 is almost non-existent.

 The graph in Fig.~\ref{fig:xtOrigin}g summarizes the strengths and phases of \xt\ between qubit 0 and 1,2, and 6.

% JAN: DO NOT REMOVE
% /xtalkmitigation/2024_paper/$  ./figF_exp234.py
%=================================================
\begin{figure}[htbp]
\centering
\includegraphics[width=0.99\linewidth]{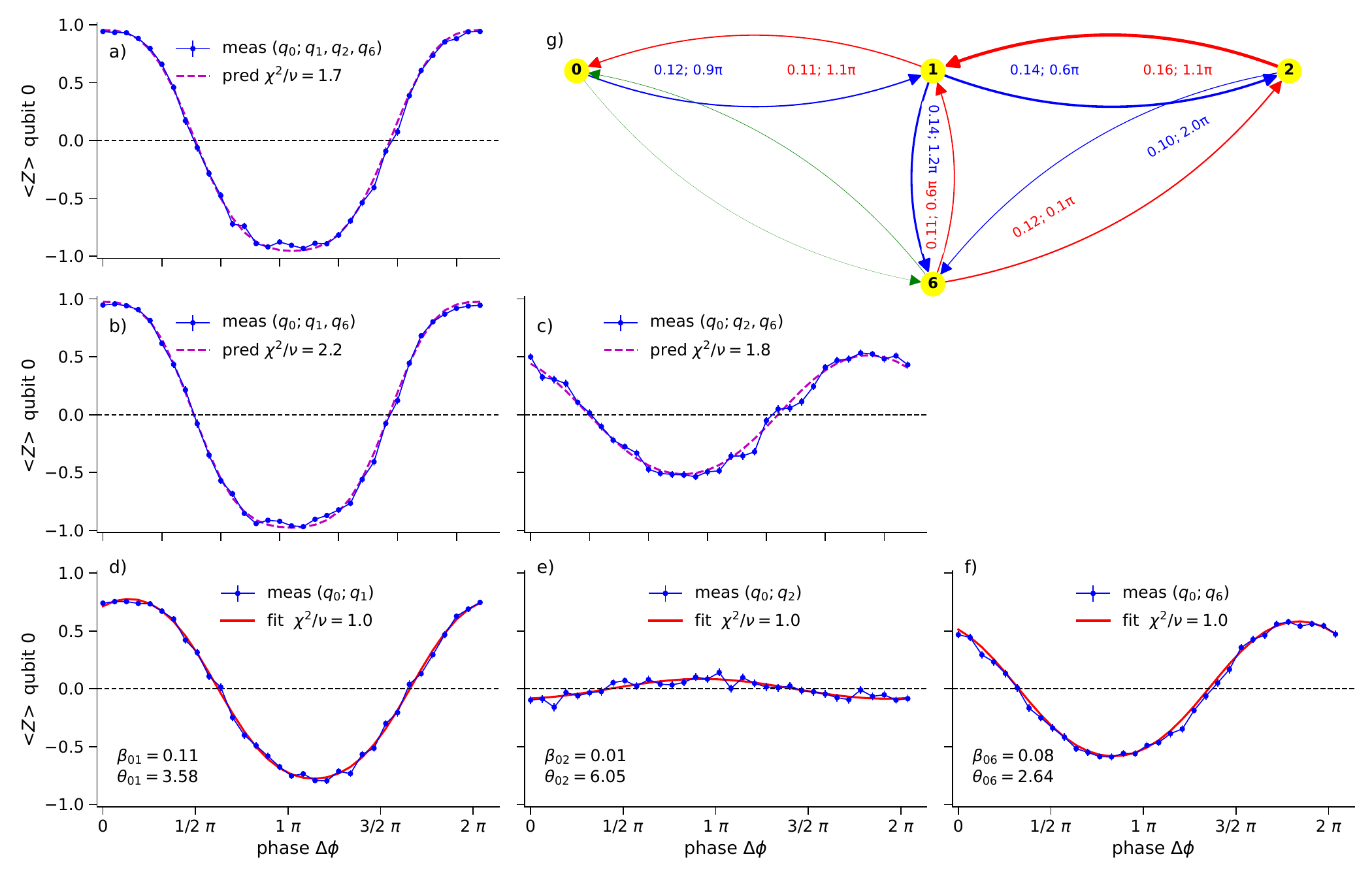}
   \caption{Accumulation of \xt\ for qubit 0 when simultaneously driving 1,2, and 6.  Panels a-c) shows predictions, d-e) shows fits, g)  graph summarizing \xt\ learned from pair-wise measurements. See the main text for the details.
}
\label{fig:xtOrigin}
\end{figure}

\end{widetext} 